\documentclass[american,aps,pra,reprint, superscriptaddress]{revtex4-1}
\usepackage[unicode=true,pdfusetitle, bookmarks=true,bookmarksnumbered=false,bookmarksopen=false, breaklinks=false,pdfborder={0 0 0},backref=false,colorlinks=false] {hyperref}
\hypersetup{ colorlinks,linkcolor=myurlcolor,citecolor=myurlcolor,urlcolor=myurlcolor}
\usepackage{graphics,epstopdf,graphicx, amsthm, amsmath, amssymb, times, txfonts, braket, colortbl, color, bm}
\usepackage[up]{subfigure}
\definecolor{myurlcolor}{rgb}{0,0,0.7}

\newcommand{\proj}[1]{| #1\rangle\!\langle #1 |}
\DeclareMathOperator{\trace}{Tr}
\newcommand{\Ptr}[2]{\trace_{#1}\Pa{#2}}
\newcommand{\Tr}[1]{\Ptr{}{#1}}
\newcommand{\Innerm}[3]{\left\langle #1 \left| #2 \right| #3 \right\rangle}
\newcommand{\Pa}[1]{\left(#1\right)}
\newcommand{\Br}[1]{\left[#1\right]}
\def\real{\mathbb{R}}

\theoremstyle{plain}
\newtheorem{thm}{\protect\theoremname}

\newtheorem{lem}[thm]{Lemma}
\providecommand{\theoremname}{Theorem}

\newcommand*{\myproofname}{Proof}
\newenvironment{mproof}[1][\myproofname]{\begin{proof}[#1]}{\end{proof}}

\newcommand{\tinyspace}{\mspace{1mu}}
\newcommand{\abs}[1]{\left\lvert\tinyspace #1 \tinyspace\right\rvert}

\def \dif {\mathrm{d}}

\makeatother

\def\I{\mathbb{I}}
\def \dif {\mathrm{d}}

\def\rS{\mathrm{S}}
\newcommand{\norm}[1]{\left\lVert\tinyspace #1 \tinyspace\right\rVert}

\begin{document}

 \author{Uttam Singh}
 \email{uttamsingh@hri.res.in}
 \affiliation{Harish-Chandra Research Institute, Allahabad, 211019, India}
 \author{Lin Zhang}
 \email{linyz@zju.edu.cn}
 \affiliation{Institute of Mathematics, Hangzhou Dianzi University, Hangzhou 310018, PR China}
 \author{Arun Kumar Pati}
 \email{akpati@hri.res.in}
 \affiliation{Harish-Chandra Research Institute, Allahabad, 211019, India}

\title{Average coherence and its typicality for random pure states}

\begin{abstract}
We investigate the generic aspects of quantum coherence guided by the concentration of measure phenomenon. We find the average relative entropy of coherence of pure quantum states sampled randomly from the uniform Haar measure and show that it is typical, i.e., the probability that the relative entropy of coherence of a randomly chosen pure state is not equal to the average relative entropy of coherence (within an arbitrarily small error) is exponentially small in the dimension of the Hilbert space. We find the dimension of a random subspace of the total Hilbert space such that all pure states that reside on it almost always have at least a fixed nonzero amount of the relative entropy of coherence that is arbitrarily close to the typical value of coherence. Further, we show, with high probability, every state (pure or mixed) in this subspace also has the coherence of formation at least equal to the same fixed nonzero amount of the typical value of coherence. Thus, the states from these random subspaces can be useful in the relevant coherence consuming tasks like catalysis in the coherence resource theory. Moreover, we calculate the expected trace distance between the diagonal part of a random pure quantum state and the maximally mixed state, and find that it does not approach to zero in the asymptotic limit. This establishes that randomly chosen pure states are not typically maximally coherent (within an arbitrarily small error). Additionally, we find the lower bound on the relative entropy of coherence for the set of pure states whose diagonal parts are at a fixed most probable distance from the maximally mixed state.
\end{abstract}

\maketitle

\section{Introduction}
\label{intro}
Random pure quantum states offer new insights for various phenomena in quantum physics and quantum information theory by exploiting the strong mathematical tools of probability theory and random matrix theory \cite{Collins2015}. These states play a fundamental role in providing a satisfactory explanation to the postulate of \emph{equal a priori probability} of statistical physics \cite{Popescu2006, Goldstein2006}. Moreover,  various properties of complex quantum systems become typical for these states allowing to infer general structures on the set of states on the Hilbert space \cite{Hayden2004, Hayden2006, Collins2015}. In particular, the entanglement properties of pure bipartite quantum states sampled from the uniform Haar measure have been studied extensively \cite{Page1993, Foong1994, Jorge1995, Sen1996, Malacarne2002, Hayden2004, Hayden2006, Datta2010, Hamma2012, Dahlsten2014, ZhangB2015, Nakata2015}. It has been shown that the overwhelming majority of random pure quantum states sampled from the uniform Haar measure are extremely close to the maximally entangled  state \cite{Hayden2006} which seems very counterintuitive. Notably, L\'evy's lemma and in general, the concentration of measure phenomenon, used in proving the above result paved the way to construct counterexamples to the conjecture of the additivity of minimum output entropy \cite{Werner2002, Hayden2008, Hastings2009} among other important implications \cite{Ledoux2005}. Also, the physical relevance of generic entanglement has been established by showing that it can be generated efficiently \cite{Oliveira2007}.

In recent years, quantum coherence has been deemed important in a wide spectrum of physical situations including quantum thermodynamics \cite{Aspuru13, Horodecki2013, Skrzypczyk2014, Narasimhachar2015, Brandao2015, Rudolph214, Rudolph114, Gardas2015, Avijit2015, Goold2015} and quantum biology \cite{Plenio2008, Aspuru2009, Lloyd2011, Li2012, Huelga13, Levi14}. This has led to the development of resource theories of coherence \cite{Gour2008, Marvian14, Baumgratz2014} adapting the well established notions of the entanglement resource theory \cite{HorodeckiRMP09}. Since then, these theories have steered various explorations of the coherence properties of quantum systems \cite{Girolami14, Bromley2015, Alex15, Xi2015, Winter2015, Fan15, Pinto2015, Du2015, Yao2015, Killoran2015, ZhangA2015, Uttam2015, UttamA2015, Cheng2015, Mondal2015, MondalA2015, Kumar2015, Mani2015, Bu2015}. Also, the possible connections between the coherence resource theories and that of entanglement have been explored \cite{Alex15, Chitambar2015, StreltsovA2015, ChitambarA2015}.
%
However, coherence properties of random pure states of single quantum systems have not been studied in great detail.
In this work, we find the behavior of the quantum coherence for a system in a pure quantum state chosen randomly from the uniform Haar measure. We show that for higher
dimensional systems the coherence behaves generically, i.e., most of the systems in these random pure states posses almost the same amount of coherence. We demonstrate that the generic nature of coherence of these states holds for various measures of coherence such as the relative entropy of coherence \cite{Baumgratz2014} which is also equal to the distillable coherence, the coherence of formation \cite{Winter2015} and the $l_1$ norm of coherence \cite{Baumgratz2014}. In these situations the coherence is solely determined by a few generic parameters that appear in the ``concentration of measure phenomenon'', such as the dimension of the Hilbert space. We find a large concentrated subspace of the full Hilbert space with the property that the relative entropy of coherence \cite{Baumgratz2014} of every pure state in this subspace is almost always lower bounded by a fixed number that is very close to the typical value of coherence. Moreover, for all the states (pure or mixed) in this subspace, the coherence of formation \cite{Winter2015} is also lower bounded by the same fixed number. These subspaces are of immense importance in situations where quantum coherence is a useful resource as they guarantee a lower bound on the amount of coherence that may be used. An important example, that consumes coherence, is the catalysis of coherence \cite{Aberg14} which allows the state transformations that are otherwise forbidden (as may be required in work extraction protocols in quantum thermodynamics) within the allowed set of operations. Furthermore, we find that most of the pure states sampled randomly from the Haar measure are not typically maximally coherent. 
This is in sharp contrast to the fact that most of the bipartite pure state sampled randomly from the Haar measure are typically maximally entangled \cite{Hayden2006}. Since the quantum coherence quantifies the wave nature of a particle \cite{Bera2015, Bagan2015}, one
may ask how `wavy' is a quantum particle if the state of the particle is chosen at random from the uniform Haar measure? Our result shows that the `typical wave nature' of a quantum particle such as a qudit is directly
related to $d$-th harmonic number.

The paper is organized as follows. We start with a discussion on random pure quantum states, measures of coherence, the concentration of measure phenomenon and a few other preliminaries in Sec \ref{sec:rand-state}. In Sec \ref{sec:avg-rel-coh}, we calculate the average relative entropy of coherence for random pure states sampled from the uniform Haar distribution, establish the typicality of the obtained average amount of coherence, and find the dimension of the subspace of the total Hilbert space with the property that all pure states in this
subspace have at least a fixed amount of relative entropy of coherence as well as coherence of formation. We then present our results on expected classical purity, its typicality and the upper bound on the $l_1$ norm of coherence in Sec \ref{sec:avg-coh}. Subsequently, in Sec \ref{sec:max-coh}, we establish that most of the randomly sampled pure states are not typically maximally coherent. Finally, we conclude in Sec \ref{sec:conclusion} with an overview on the implications of the results presented in the paper.
\section{Random pure quantum states, measures of coherence and concentration of measure phenomenon}\label{sec:rand-state}
\smallskip
\noindent {\it  Random pure states:}
It only makes sense to talk about random quantum states after we have fixed a measure $\mu$ on the set of quantum states. Having fixed a measure $\mu$ on the set of quantum states one can calculate the desired averages over all states with respect to this measure. Here we are interested in the set of pure quantum states. For a $d$-dimensional Hilbert space $\mathcal{H}$, the set of pure states is identified as complex projective space $\mathbb{C}P^{d-1}$. On this space there exists a unique natural measure $\dif(\psi)$, induced by the uniform Haar measure $\dif\mu(U)$ on the unitary group $\mathrm{U}(d)$ \cite{Wootters1990, Zyczkowski1994, Zyczkowski2001, Bengtsson2008, Aubrun2014}. This amounts to saying that any random pure state $\ket{\psi}$ is generated equivalently by applying a random unitary matrix $U\in \mathrm{U}(d)$ on a fixed pure state $\ket{\psi_0}$, i.e., $\ket{\psi} = U\ket{\psi_0}$. Now one can define the average value of some function $g$ of pure state as follows:
\begin{align}
\mathbb{E}_\psi g(\psi) :=\int \dif(\psi) ~g(\psi) = \int_{\mathrm{U}(d)} \dif\mu(U)~ g(U\psi_0).\nonumber
\end{align}
In what follows, by random pure states we mean the states generated by applying random Haar distributed unitaries on some fixed pure state and all the averages are with respect to the Haar measure.

\smallskip
\noindent {\it  Measures of coherence:}
The measures of coherence that we consider in this work are the $l_1$ norm of coherence, the relative entropy of coherence and the coherence of formation. For a density matrix $\rho$ of dimension $d$ and a fixed reference basis $\{\ket{i}\}$, the $l_1$ norm of coherence $C_{l_1}(\rho)$ \cite{Baumgratz2014} is defined as
\begin{align}
C_{l_1}(\rho) = \sum_{\substack{{i,j=1}\\{i\neq j}}}^d |\bra{i}\rho\ket{j}|.
\end{align}
The relative entropy of coherence $C_r(\rho)$ \cite{Baumgratz2014} is defined as
\begin{align}
 C_r(\rho) = S(\rho_D) - S(\rho),
\end{align}
where $\rho_D$ is the diagonal part of the density matrix $\rho$ in the fixed reference basis and $S$ is the von Neumann entropy defined as $\rS(\rho) = -\Tr{\rho\ln\rho}$. Here and in the rest of the paper, all the logarithms are taken with respect to the base $e$. The coherence of formation $C_f(\rho)$ \cite{Winter2015} is defined as
\begin{align}
C_f(\rho) = \min_{\{p_a,\ket{\psi_a}\bra{\psi_a}\}}\sum_{a}p_a S(\rho_D(\psi_a)),
\end{align}
where $\rho_D(\psi_a)$ is the diagonal part of the pure state $\ket{\psi_a}$, $\rho = \sum_ap_a\ket{\psi_a}\bra{\psi_a}$ and minimum is taken over all such decompositions of $\rho$. We emphasize that here we consider an intrinsically basis dependent notion of coherence applicable to finite dimensional systems. It may be remarked that the above notion of coherence has attracted a great deal of interest recently, although it is not widely accepted by all. There are other notions of coherence such as those based on the resource theory of asymmetry \cite{Gour2008, Marvian14} and the optical coherence in quantum optics \cite{Mandel95, Vogel2014}.


\smallskip
\noindent {\it  Concentration of measure phenomenon:}
For many functions defined over a vector space, the overwhelming majority of vectors take a value of the function very close to the average value as the dimension of the vector space goes to infinity.
This observation, collectively, is referred to as the concentration of measure phenomenon. Here we show that several measures of coherence have this property. Let us consider a simple example to demonstrate the concentration of measure phenomenon. Consider the $k$-sphere $\mathbb{S}^{k}$ in $\mathbb{R}^{k+1}$ with $k$ being very large. A direct calculation yields that the uniform measure $\mu$ on $\mathbb{S}^{k}$ is almost concentrated around every equator when $k$ is large. Similarly, an explicit calculation \cite{Ledoux2005} of the measure of spherical caps implies that given any measurable set $\mathcal{S}$ with $\mu(\mathcal{S})\geq 1/2$, for every $r>0$, $\mu(\mathcal{S}_r)\geq 1- \exp\{{(k-1)r^2/2}\}$ where $\mathcal{S}_r = \{x\in \mathbb{S}^k: d(x,\mathcal{S})<r\}$ and $d(x,y)$ is the Euclidean distance on $\mathbb{R}^{k+1}$. This is one of the first quantitative instances of the concentration of measure phenomenon. For Lipschitz continuous functions on the sphere,  L\'evy's lemma is the rigorous statement about the concentration of measure phenomenon \cite{Ledoux2005}. Let us first define the Lipschitz continuous functions.

\smallskip
\noindent {\it Lipschitz continuous function and Lipschitz constant:}
Suppose $(M, d_1)$ and $(N, d_2)$ are metric spaces and $F : M \rightarrow N$. If there exists $\eta \in \mathbb{R}^+$ such that $d_2(F(x),F(y)) \leq \eta d_1(x, y)$ for all $x, y \in M$, then $F$
is called a Lipschitz continuous function on $M$ with the Lipschitz constant $\eta$. Every real number larger than $\eta$ is also a Lipschitz constant for $F$ \cite{Searcoid2007}. Next, we introduce a form of L\'evy's lemma that will be the key ingredient in our paper.


\smallskip
\noindent {\it L\'evy's Lemma (see \cite{Ledoux2005} and \cite{Hayden2006}):} Consider a sequence $F=\{F_k : \mathbb{S}^{k} \rightarrow \mathbb{R}\}_k$ of Lipschitz continuous functions from the $k$-sphere to the real line with each function $F_k$ having the same Lipschitz constant $\eta$ that is independent of $k$ (with respect to the Euclidean norm).
Let a point $X \in \mathbb{S}^{k}$ be chosen uniformly at random. Then, for all $\epsilon>0$ and $k$,
\begin{align}
\label{Levy-lemma}
\mathrm{Pr} \left\{|F_k(X) - \mathbb{E} (F_k)|  > \epsilon \right\} \leq 2\exp\left( -\frac{(k+1) \epsilon^2}{ 9\pi^3 \eta^2\ln 2 } \right).
\end{align}
Here $\mathbb{E}(F_k)$ is the mean value of $F_k$.
It is insightful to consider $\epsilon = r^{-1/4}$ in Eq. (\ref{Levy-lemma}).
With this choice, the bound on the right hand side decreases exponentially
as $\exp{(-\sqrt{r})}$ while the bound on the left hand side decreases like $r^{-1/4}$, making it clear that the probability of being non-typical decreases much faster and hence ``essentially zero" for large $r$. Note that the average over the Haar distributed $d$-dimensional pure states is equivalent to the average over the $k$-sphere with $k=2d-1$.

At various places in our work, we use the trace norm and the Euclidean norm for matrices: (1) the trace norm of a matrix $A$, denoted by $||A||_1$, is defined as $||A||_1 = \mathrm{Tr}\sqrt{A^\dag A}$ , where $\dag$ is the Hermitian conjugate. (2) the Euclidean norm of a matrix $A$, denoted by $||A||_2$, is defined as $\sqrt{\mathrm{Tr}(A^\dag A)}$. The trace distance between two density matrices $\rho$ and $\sigma$ is defined as $||\rho-\sigma||_1$ \cite{Wilde13}. Notice that we follow a definition of trace distance without a factor of half in front of the trace norm. Finally, for proving the existence of concentrated subspaces with fixed amount of coherence we need the notion of {\it small nets} \cite{Hayden2004}.

\smallskip
\noindent{\it Existence of small nets:} It is known \cite{Hayden2004} that given a Hilbert space $\mathcal{H}$ of dimension $d$ and $0 < \epsilon_0 < 1$, there exists a set $\mathcal{N}$ of pure states in $\mathcal{H}$ with $|\mathcal{N}|\leq (5/\epsilon_0)^{2d}$, such that for every pure state $\ket{\psi}\in \mathcal{H}$ there exists $\ket{\tilde\psi} \in \mathcal{N}$ with $||~\ket{\psi} - \ket{\tilde\psi}||_2 \leq \epsilon_0/2$. Such a set is called as an $\epsilon_0$-net.

We emphasize here that all the main results presented below are based on L\'evy's lemma and hence are of probabilistic nature. The method to demonstrate the typical properties is always to prove that the opposite is an unlikely event.

\section{Average relative entropy of coherence and its typicality for random pure states}\label{sec:avg-rel-coh}
To show the typicality of coherence of random pure quantum states we first find the average relative entropy of coherence for a random pure state, where average is taken over the uniform Haar measure, and then apply L\'evy's lemma to show the concentration effect for quantum coherence. Now consider a pure state $\ket{\psi}$ and denote by $\rho_D(\psi)$ the diagonal part of $\ket{\psi}$ in the fixed reference basis $\{\ket{i}\}$, i.e., $\rho_D\left (\psi \right) = \sum_{i=1}^{d} |\bra{i}\psi\rangle|^2
\Pi_i$, where $\Pi_i= \ket{i}\bra{i}$. The relative entropy of coherence of the state $\ket{\psi}$ in the fixed reference basis $\{\ket{i}\}$ is $  C_{r}\left (\psi \right) = S(\rho_D\left (\psi\right)) = -\sum_{i=1}^{d} |\bra{i}\psi\rangle|^2 \ln |\bra{i}\psi\rangle|^2$. If we draw pure states $\ket{\psi}$ from the uniform Haar measure then the expected value of the relative entropy of coherence is given by
\begin{align}
\mathbb{E}_{\psi}C_{r}\left (\psi \right) := -\sum_{i=1}^{d}\int \dif(\psi)~|\bra{i}\psi\rangle|^2 \ln |\bra{i}\psi\rangle|^2.
\end{align}
As discussed earlier, we can take $\ket{\psi} = U\ket{1}$ where $U$ is sampled from the Haar distribution and $\ket{1}$ is a fixed state. This allows us to rewrite the above equation as
\begin{align}
\mathbb{E}_{\psi}C_{r}\left (\psi \right)  = -\sum_{i=1}^{d}\int \dif\mu(U)~ |\bra{i}U|1\rangle|^2 \ln |\bra{i}U|1\rangle|^2.
\end{align}
Since the Haar measure is invariant under the left translation, we have
\begin{align}
\mathbb{E}_{\psi}C_{r}\left (\psi \right) = -d\int \dif\mu(U)~ |U_{11}|^2 \ln |U_{11}|^2,
\end{align}
where $U_{11} = \bra{1}U|1\rangle$. Note that all entries $U_{ij}$ of a Haar unitary $U$ has the same distribution \cite{Hiai2000}: $\frac{d-1}{\pi}(1-r^2)^{d-2}r\dif r\dif\theta$, where $r=|U_{ij}|\in[0,1]$ and $\theta\in[0,2\pi]$. We remark here that the distribution of each entry $U_{ij}=re^{\mathrm{i}\theta}$ is just the joint distribution of $r$ and $\theta$. The distribution of $|U_{11}|^2 $ is given by $(d-1)(1-r)^{d-2} \dif r$, where $0\leq r\leq 1$. Now,  we have
\begin{align} \label{Eq:rel}
\mathbb{E}_{\psi}C_{r}\left (\psi \right)
 &= -d(d-1)\int_{0}^{1} r(1-r)^{d-2}\ln r ~\dif r\nonumber\\
 &=  -d(d-1) \frac{\partial B(\alpha,\beta)}{\partial \alpha}\Big|_{(\alpha, \beta) = (2, d-1)},
\end{align}
where $B(\alpha,\beta) $ is the Beta function, defined as 
\begin{align}
B(\alpha,\beta) := \int_{0}^{1} r^{\alpha-1} (1-r)^{\beta-1} \dif r = \frac{\Gamma{(\alpha)} \Gamma{(\beta)}}{ \Gamma{(\alpha + \beta)} }. 
\end{align}
Note that $\partial B(\alpha,\beta)/\partial \alpha = (\Psi(\alpha)- \Psi(\alpha+\beta))B(\alpha,\beta)$, where $\Psi(z) :=\Gamma' (z)/\Gamma(z)$ and $\Gamma(z)=\int^\infty_0 x^{z-1}e^{-x}\dif x$, with $\mathrm{Re}(z)>0$, is the Gamma function. In particular, for natural number $n$, $\Psi(n) = \sum_{k=1}^{n-1}1/k-\gamma$ with $\gamma\approx 0.57721$ being the Euler constant. Therefore, we get $\partial B(\alpha,\beta)/\partial \alpha|_{(\alpha, \beta) = (2, d-1)} = -(d(d-1))^{-1} \sum_{k=2}^{d}1/k$. Using this in Eq. (\ref{Eq:rel}), we have 
$\mathbb{E}_{\psi}C_{r}\left (\psi \right) = \sum_{k=2}^{d}\frac{1}{k}$.
Thus, a $d$-dimensional random pure state has $H_d -1$ amount of average relative entropy of coherence, where $H_d=\sum_{k=1}^{d}1/k$ is the $d$-th harmonic number.
Now we are ready to discuss the concentration of measure phenomenon for quantum coherence.

\begin{figure}
\subfigure[~$d$ = 20]{
\includegraphics[width=40 mm]{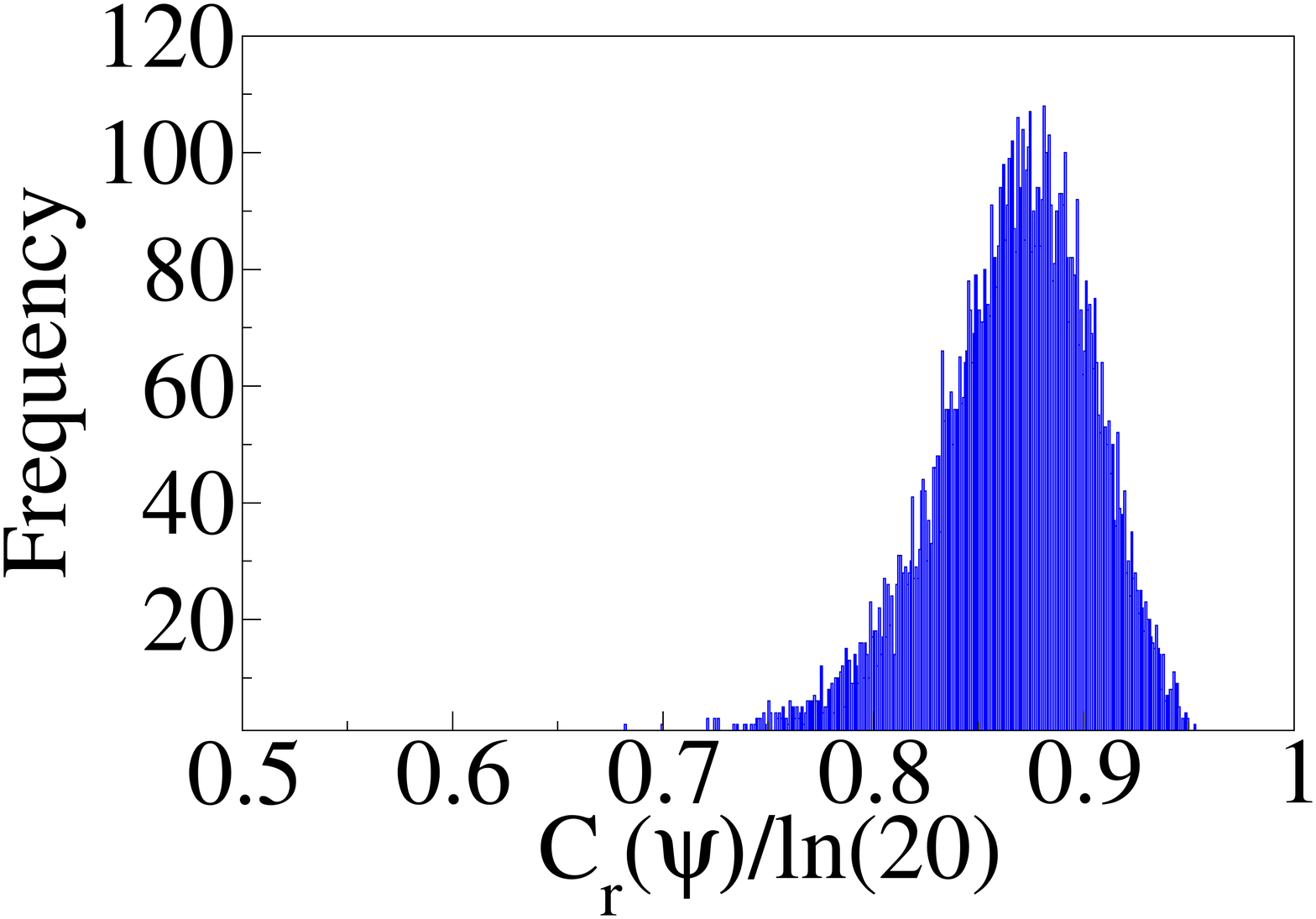}}
\subfigure[~$d$ = 30]{
\includegraphics[width=40 mm]{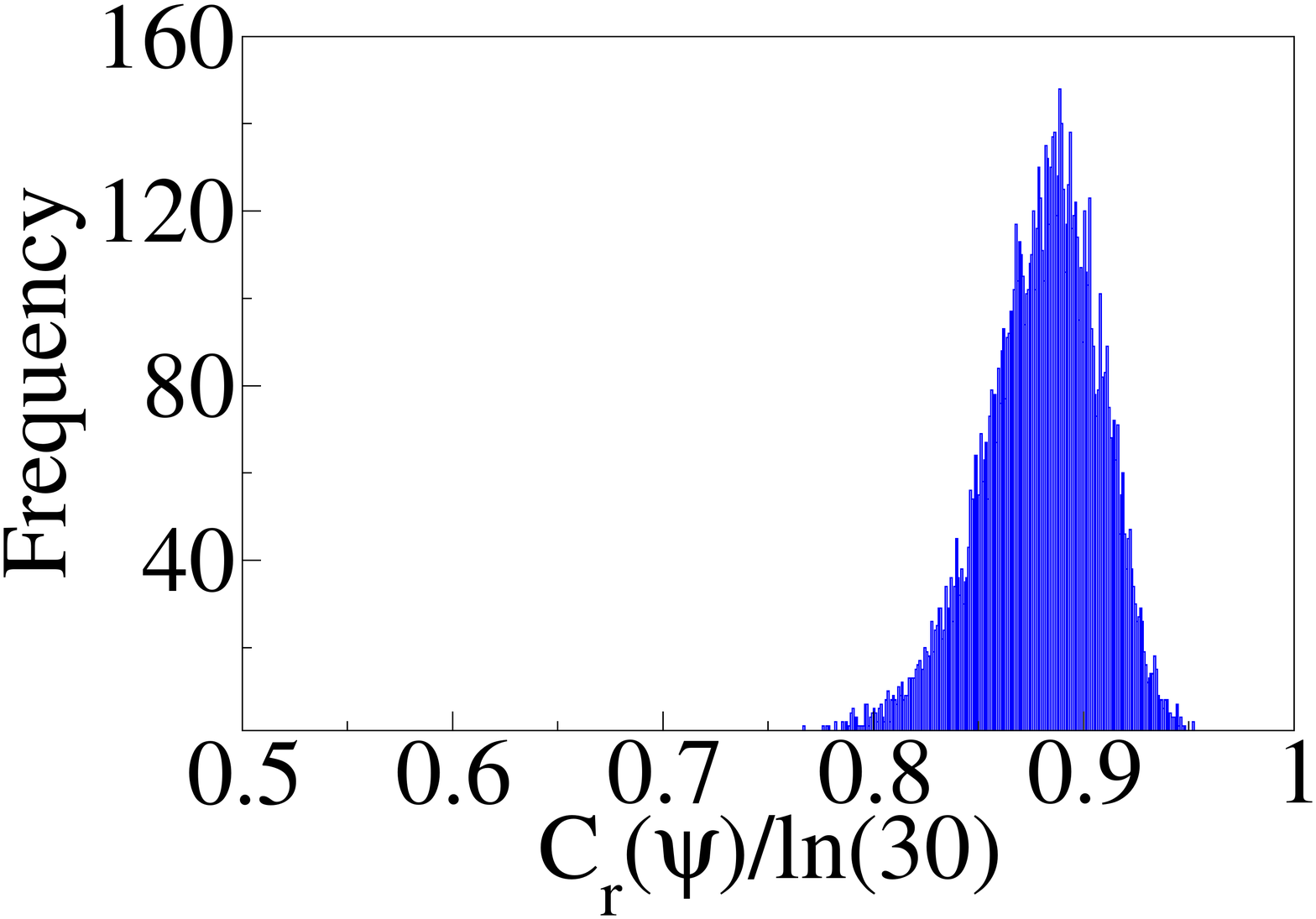}}\\
\subfigure[~$d$ = 40]{
\includegraphics[width=40 mm]{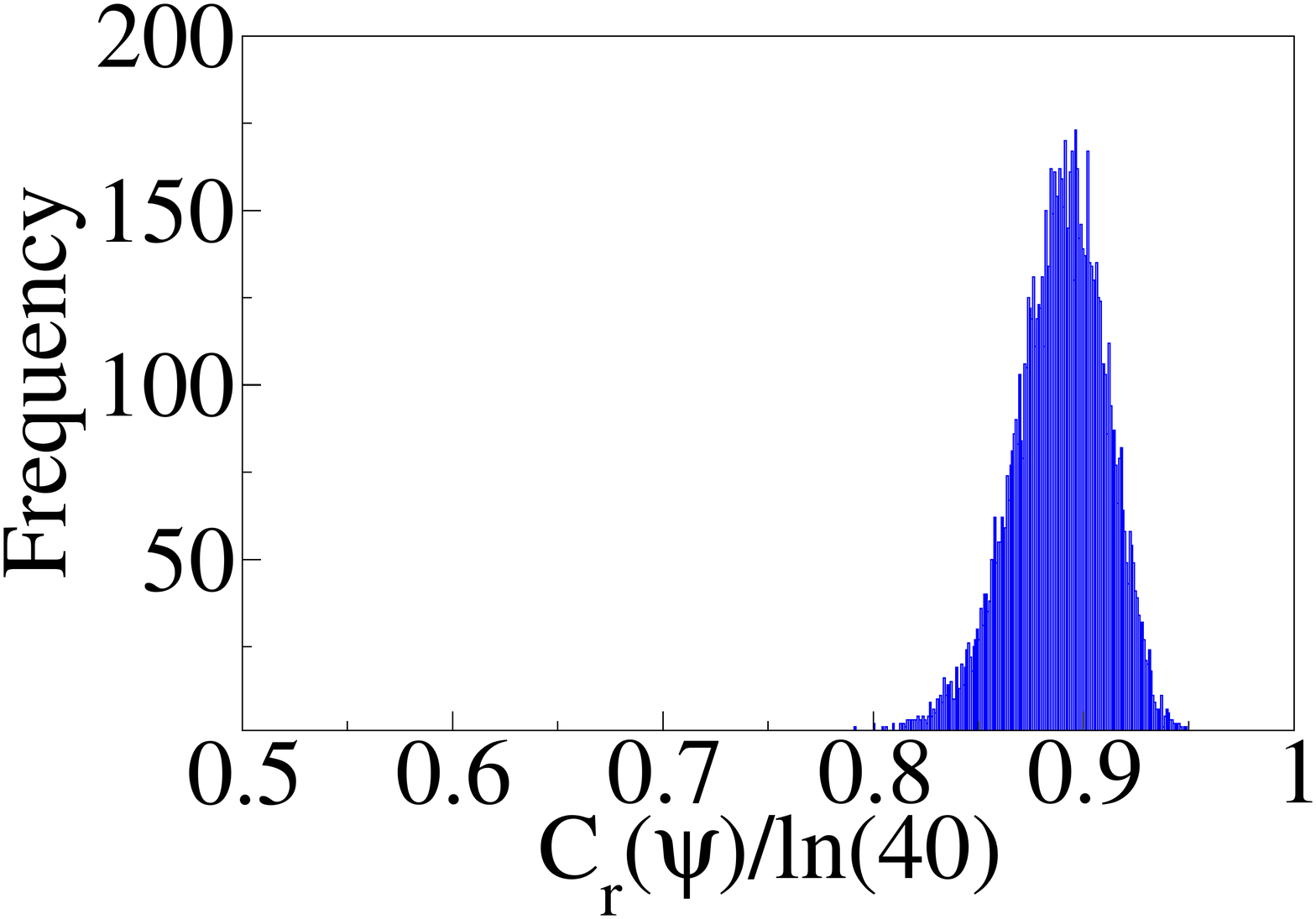}}
\subfigure[~$d$ = 500]{
\includegraphics[width=40 mm]{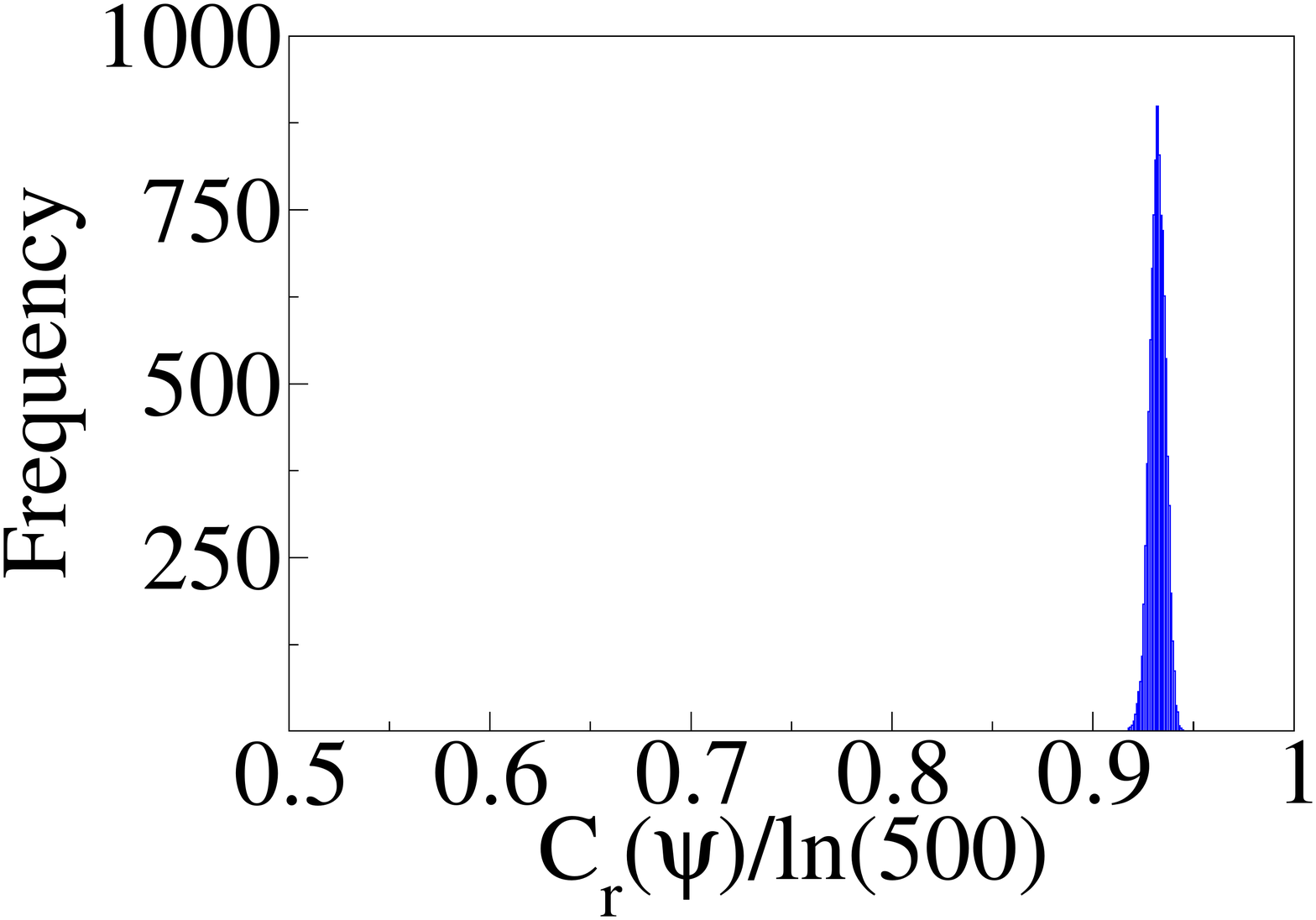}}
\caption{(Color online) The frequency plot showing the (scaled) relative entropy of coherence $C_r(\psi)/\ln d$ for the Haar distributed random pure states for dimensions $d=20, 30, 40 \mbox{~and~} 500$. Here, both the axes are dimensionless. We have $\mathbb{E}_\psi C_r(\psi)/\ln 20\approx 0.87$, $\mathbb{E}_\psi C_r(\psi)/\ln 30\approx 0.88$, $\mathbb{E}_\psi C_r(\psi)/\ln 40\approx 0.89$ and $\mathbb{E}_\psi C_r(\psi)/\ln 500\approx 0.93$. 
The plot shows that the (scaled) relative entropy of coherence is indeed very close to the average value $\sum_{k=2}^{d}1/k$.
As we increase the dimension, the figure shows that more and more states have coherence close to the average value and the variances approach to zero.}
\label{fig}
\end{figure}

\smallskip{}
\begin{thm}[\textbf{Concentration of the relative entropy of coherence}]
\label{th:con-ent}
Let $\ket{\psi}$ be a random pure state on a $d$-dimensional Hilbert space $\mathcal{H}$ with $d\geq 3$. Then
for all $\epsilon > 0$
\begin{align}
\label{Eq:conc}
\mathrm{Pr} &\left\{ \left|C_r(\psi) -\left(H_d-1\right)\right|> \epsilon \right\} \leq 2\exp\left( -\frac{d \epsilon^2}{36 \pi^3 \ln 2 (\ln d)^2} \right),
\end{align}
where $H_d=\sum_{k=1}^{d}1/k$ is the $d$-th harmonic number.
\end{thm}

\smallskip{}
\begin{mproof}
We will apply L\'evy's lemma, Eq. (\ref{Levy-lemma}) for averages, to prove the theorem.
Consider the map $F: \ket{\psi} \rightarrow F(\psi):= S(\rho_D\left (\psi \right)) = C_r(\psi)$. We have already shown that $\mathbb{E}_{\psi}F =H_d-1$. We prove the theorem by identifying $k$ with $2d-1$ in L\'evy's lemma (Eq. (\ref{Levy-lemma})). We just need the Lipschitz constant $\eta$ for the function $F$ such that $|F(\psi) - F(\phi)| \leq \eta || \ket{\psi}-\ket{\phi} ||_2$. Let $\ket{\psi} = \sum_{i=1}^{d} \psi_i \ket{i}$ and therefore, $\rho_D\left( \psi\right) = \sum_{i=1}^{d}p_i(\psi) \ket{i}\bra{i}$ with $p_i(\psi)=|\psi_i|^2$. Now, $ F(\psi) = -\sum_{i=1}^{d} p_i(\psi) \ln p_i(\psi)$. The Lipschitz constant for $F$ can be bounded as follows:
\begin{align}
\label{eq:lip-one}
\eta^2 := \sup_{\bra{\psi}\psi\rangle\leq 1} \nabla F \cdot \nabla F
&= 4\sum_{i=1}^{d}p_i(\psi)\left[ 1 + \ln p_i(\psi) \right]^2\nonumber\\
&\leq 4 \left(1+ \sum_{i=1}^{d} p_i(\psi)(\ln p_i(\psi))^2\right)\nonumber\\
&\leq 4 \left(1+ (\ln d)^2\right)\leq 8(\ln d)^2,
\end{align}
where the last inequality is true for $d\geq 3$. Therefore, $\eta \leq \sqrt{8}\ln d$ for $d\geq 3$. By definition, any upper bound on the Lipschitz constant can also serve as a valid Lipschitz constant, therefore, we can take $\eta=\sqrt{8}\ln d$ for $d\geq 3$.
This concludes the proof of the theorem. 
\end{mproof}
The inequality (\ref{Eq:conc}) means that for large $d$, the number of pure states with the relative entropy of coherence not very close to $H_d-1$ are exponentially small, or in other words, most pure states
chosen randomly have $H_d-1$ amount of relative entropy of coherence to within an arbitrarily small error. This is the concentration of relative entropy of coherence around its expected value (the typical value of the relative entropy of coherence). 
Further, as quantum coherence is a quantifier of the wave nature of a quantum particle \cite{Bera2015,Bagan2015}, Theorem \ref{th:con-ent}  has a nice physical meaning and it quantifies the `typical wave nature' of a random pure state.
Fig. \ref{fig} plots the relative entropy of coherence for numerically generated Haar distributed random pure quantum states and shows that indeed most of the states have coherences close to the expected value. 
%

Having established the concentration of relative entropy of coherence, it is of great practical importance to delineate the largest subspace of the total Hilbert space such that all the pure states in this subspace have a fixed nonzero amount of coherence. Specifically, we find a large subspace of the total Hilbert space such that the amount of the relative entropy of coherence for every pure state in this subspace can be bounded from below almost always by a number that is arbitrarily close to the typical value of coherence. The following theorem formalizes this.

\begin{thm}[\textbf{Coherent subspaces}]
\label{thm:coh-space}
Let $\mathcal{H}$ be a Hilbert space of dimension $d\geq 3$ of a quantum system. Then, for any positive $\epsilon < \ln d$, there exists a subspace $\mathcal{S}\subset \mathcal{H}$ of dimension
\begin{align}
\label{Eq:dim} s=\left\lfloor d K \left(\frac{\epsilon}{\ln d}\right)^{2.5}\right\rfloor
\end{align}
such that all pure states $\ket{\psi} \in \mathcal{S}$ almost always satisfy $C_r(\psi) \geq H_d-1- \epsilon$.
$K$ may be chosen to be $1/16461$. Here $\lfloor \rfloor$ denote the floor function.
\end{thm}
\smallskip{}
\begin{mproof}
Here we follow the strategy of Ref. \cite{Hayden2006} which is based on the construction of nets to prove the theorem.
Let $\mathcal{S}$ be a random subspace of $\mathcal{H}$ of dimension $s$. Let $\mathcal{N}_\mathcal{S}$ be an $\epsilon_0$-net for states on $\mathcal{S}$, for $\epsilon_0=\epsilon/(\sqrt{8} \ln d)$. By definition, we have $|\mathcal{N}_\mathcal{S}|
\leq (5/\epsilon_0)^{2s}$. Note that $\mathcal{S}$ may be thought of as $U\mathcal{S}_0$, with a fixed $\mathcal{S}_0$ and a unitary $U$ distributed according to the Haar measure. We can fix the net $\mathcal{N}_{\mathcal{S}_0}$ on $\mathcal{S}_0$ and let $\mathcal{N}_{\mathcal{S}}=U\mathcal{N}_{\mathcal{S}_0}$. This is a natural way to choose a random subspace. Now, given $\ket{\psi}\in \mathcal{S}$, we can choose $\ket{\tilde\psi} \in \mathcal{N}_\mathcal{S}$ such that $||~\ket{\psi} - \ket{\tilde\psi}||_2 \leq \epsilon_0/2$. Note that $C_r(\psi)$ is a Lipschitz continuous function with the Lipschitz constant $\eta=\sqrt{8} \ln d$. From definition of the Lipschitz function and $\epsilon_0$-net, we have
\begin{align}
|C_r(\psi) - C_r(\tilde\psi)| \leq \eta ||~\ket{\psi} - \ket{\tilde\psi}||_2 \leq \eta \epsilon_0/2 = \epsilon/2.\nonumber
\end{align}
Define $\mathbb{P} = \mathrm{Pr} \left\{\inf_{\ket{\psi}\in \mathcal{S}}C_r(\psi) < H_d-1 - \epsilon \right\}$. Now, we have
\begin{align}
\mathbb{P} &\leq \mathrm{Pr} \left\{\min_{\ket{\tilde\psi}\in \mathcal{S}}C_r(\tilde\psi) < H_d-1 - \epsilon/2 \right\}\nonumber\\
&\leq |\mathcal{N}_\mathcal{S}|~\mathrm{Pr} \left\{C_r(\psi) < H_d-1 - \epsilon/2 \right\}\nonumber\\
&\leq 2\left(10\sqrt{2}\ln d/\epsilon\right)^{2s}\exp\left( -\frac{d \epsilon^2}{144 \pi^3 \ln 2 (\ln d)^2} \right),
\end{align}
where in the last line we have used our Theorem \ref{th:con-ent} and the definition of $\epsilon_0$-net. If this probability is smaller than one, a subspace with the stated properties will exist. This can be assured by choosing
\begin{align}
s < \frac{(d-1) \epsilon^2}{6190 (\ln d)^2 \ln \left(\left(10\sqrt{2}\ln d\right)/\epsilon\right)}.
\end{align}
Now, using the fact that $\ln x \leq \sqrt{x/2}$ for $x\geq 10\sqrt{2}$, we have $\ln \left((10\sqrt{2}\ln d)/\epsilon\right)\leq \sqrt{5\sqrt{2}\ln d/\epsilon}$ with $\epsilon<\ln d$. For a nontrivial dimension $s$, i.e., $s\geq 2$, we require $d\geq 32921$. Therefore, $ s = \left\lfloor \frac{d \epsilon^{2.5}}{16461(\ln d)^{2.5}}\right\rfloor$. This completes the proof of the theorem.
\end{mproof}
The theorem implies that if a subspace of dimension $s$ (which can be appropriately large), given by Eq. (\ref{Eq:dim}), of total Hilbert space is chosen at random via the Haar distribution then the relative entropy of coherence of any pure state in this subspace is almost always greater than $H_d-1 - \epsilon$, which is very close to the typical value of coherence. This follows from the fact that the probability that the chosen subspace will not have the above said property is small. Now, for any pure state $\ket{\psi}$ in $\mathcal{S}$, the relative entropy of coherence $C_r(\psi)$ is typically lower bounded by $H_d-1 - \epsilon$. Therefore, for all $\rho\in \mathcal{S}$, the coherence of formation which is defined as $
C_{f}(\rho) = \min\sum_{i} p_i S(\rho_D(\psi_i)) \mathrm{~such~that~} \rho = \sum_i p_i \ket{\psi_i}\bra{\psi_i}$  \cite{Winter2015},
is also typically lower bounded by $H_d-1 - \epsilon$, i.e.,  $C_{f}(\rho) \geq H_d-1 - \epsilon$.

\section{Average classical purity of random pure quantum states}\label{sec:avg-coh}
In this section, we calculate the average classical purity \cite{Cheng2015} of random pure quantum states and show its typicality.
It is not straightforward to find the expected value of the $l_1$ norm of coherence for random pure states. Therefore, we resort to an indirect method to obtain an upper bound on it using the expected value of classical purity. The classical purity $P(\Pi(\psi))$ of a state $\ket{\psi}$ is defined as $P(\Pi(\psi)):=\mathrm{Tr}[(\Pi(\psi))^2]$ where $\Pi:\rho\rightarrow \sum_{i}\bra{i}\rho\ket{i}\ket{i}\bra{i}$, i.e., it maps any state to its diagonal part in a fixed basis $\{\ket{i}\}$ \cite{Cheng2015}. For a pure state $\ket{\psi}$, we have $\Pi(\psi)=\rho_D(\psi)$. The expected classical purity $\mathbb{E}_\psi P(\Pi(\psi))$ can be obtained as follows. For a random pure state $\ket{\psi}$ sampled from the uniform Haar measure the expected classical purity is given by 
\begin{eqnarray}
\mathbb{E}_\psi P(\Pi(\psi)) =\int \dif(\psi) P(\Pi(\psi)) = \int_{\mathrm{U}(d)} \dif\mu(U) P(\Pi(U\psi_0)).\nonumber
\end{eqnarray}
Let $\Phi$ be a linear super-operator that transforms a random pure state $\proj{\psi}$ to $\Phi(\proj{\psi})$. The purity of the state $\Phi(\proj{\psi})$ is defined as $\mathrm{Tr}[\Phi(\proj{\psi})^2]$. Therefore, the expected purity for the states $\Phi(\proj{\psi})$ is given by
\begin{align}
\label{eq:int-step}
\mathbb{E}_\psi P(\Phi(\psi)) &= \int \dif(\psi) \mathrm{Tr}[\Phi(\proj{\psi})^2]\nonumber\\
&= \int\dif(\psi)\Tr{\Phi^\dagger\circ\Phi(\proj{\psi})\proj{\psi}}\nonumber\\
&=\Innerm{\psi_0}{\int\dif\mu(U)U^\dagger\Phi^\dagger\circ\Phi(U\proj{\psi_0}U^\dagger)U}{\psi_0},
\end{align}
where $\Phi^\dagger$ is the dual of $\Phi$ in the sense: $\Tr{Y\Phi(X)}=\Tr{\Phi^\dagger(Y)X}$ for any $X,Y$ and $\ket{\psi_0}$ is a fixed state such that $\ket{\psi} = U\ket{\psi_0}$. We use the following formula from matrix integral \cite{Zhang2014}
\begin{align}\label{eq:lin}
\int\dif\mu(U) U^\dagger \Upsilon(UXU^\dagger)U = &\frac{d\Tr{\Upsilon(\I_d)}-\Tr{\Upsilon}}{d(d^2-1)}\Tr{X}\I_d \nonumber\\
&+ \frac{d\Tr{\Upsilon}-\Tr{\Upsilon(\I_d)}}{d(d^2-1)}X,
\end{align}
where $\Tr{\Upsilon}$ is the trace of the super-operator $\Upsilon$, defined by $\Tr{\Upsilon}=\sum^d_{i,j=1}\Innerm{i}{\Upsilon(\ket{i}\bra{j})}{j}$, to simplify Eq. (\ref{eq:int-step}). Now identifying $X$ with $\proj{\psi_0}$  and $\Upsilon$ with
$\Phi^\dagger\circ\Phi$ in Eq. (\ref{eq:lin}), we get
\begin{align}
\mathbb{E}_\psi P(\Phi(\psi))&=\frac{d\Tr{\Phi^\dagger\circ\Phi(\I_d)}-\Tr{\Phi^\dagger\circ\Phi}}{d(d^2-1)}\nonumber\\
&~~~+\frac{d\Tr{\Phi^\dagger\circ\Phi}-\Tr{\Phi^\dagger\circ\Phi(\I_d)}}{d(d^2-1)}\nonumber\\
&=\frac1{d(d+1)}\Br{\Tr{\Phi^\dagger\circ\Phi(\I_d)}+\Tr{\Phi^\dagger\circ\Phi}}.\nonumber
\end{align}
Let $\Phi=\Pi$, then $\Pi^\dagger=\Pi$ and $\Pi\circ\Pi=\Pi$. Moreover, $ \Tr{\Pi^\dagger\circ\Pi(\I_d)}=d~~\text{and}~~\Tr{\Pi^\dagger\circ\Pi}=d$. The expected classical purity, therefore, is given by
\begin{eqnarray}
\mathbb{E}_\psi P(\Pi(\psi)) =\frac{2}{d+1}.
\end{eqnarray}
The following theorem establishes that the $\mathbb{E}_\psi P(\Pi(\psi))$ is a typical property
of the pure quantum states sampled from the uniform Haar distribution.

\smallskip{}
\begin{thm}[\textbf{Concentration of classical purity}]
\label{th:con-pure}
Consider a random pure state $\ket{\psi}$ in a $d$ dimensional Hilbert space. The classical purity of any pure state sampled from the Haar distribution, for all $\epsilon > 0$, satisfies
\begin{align}
\label{Eq:conc2}
\mathrm{Pr} &\left\{\left|P(\Pi(\psi)) - \frac{2}{d+1}\right|>\epsilon \right\}\leq 2\exp\left( -\frac{d \epsilon^2}{18 \pi^3\ln 2} \right).
\end{align}
\end{thm}

\smallskip{}
\begin{mproof}
We use L\'evy's lemma, Eq. (\ref{Levy-lemma}), to prove the theorem. For this we need the Lipschitz constant for the function $G:\ket{\psi}\rightarrow P(\Pi(\psi))$. Noting that $P(\Pi(\psi)) = ||\Pi(\psi)||_2^2$, we have
\begin{align}
\left| P(\Pi(\psi)) - P(\Pi(\phi))  \right|
&= \left| (||\Pi(\psi)||_2 - ||\Pi(\phi)||_2)(||\Pi(\psi)||_2+||\Pi(\phi)||_2)\right|\nonumber\\
&\leq ||\Pi(\psi)-\Pi(\phi)||_2 \left(||\Pi(\psi)||_2+||\Pi(\phi)||_2\right)\nonumber\\
&\leq 2||\Pi(\psi)-\Pi(\phi)||_2\nonumber\\
&\leq 2||\ket{\psi}-\ket{\phi}||_2.
\end{align}
Here in the second line we have used the reverse triangle inequality. In the third line we have used the fact that the purity is upper bounded by $1$ and in the last line, we have used the monotonicity of the Euclidean norm under the map $\Pi$. Therefore, the Lipschitz constant for the function $G:\ket{\psi}\rightarrow P(\Pi(\psi))$ can be chosen to be $2$. Now applying L\'evy's lemma to the function $G$ and noting $k=2d-1$, the proof of the theorem follows.
\end{mproof}
Now we exploit the relation between the $l_1$ norm of coherence and the classical purity \cite{Cheng2015} to get an upper bound on the
$l_1$ norm of coherence \footnote{Note that some of the results on the average $l_1$ norm of coherence were mentioned in Ref. \cite{Cheng2015}.}, which is
\begin{align}
\label{purity-bound}
C_{l_1}(\psi) \leq \sqrt{d(d-1) \left[1 - P(\Pi(\psi)) \right]}.
\end{align}
Since the classical purity of a random pure state is concentrated on its expected value $\mathbb{E}_\psi P(\Pi(\psi)) = 2/(d+1)$ (see Theorem \ref{th:con-pure}), one may replace $P(\Pi(\psi))$ by $2/(d+1)$ in Eq. (\ref{purity-bound}) to get an upper bound on the $l_1$ norm of coherence which depends only on the dimension of the Hilbert space. Thus, $C_{l_1}(\psi) \leq \sqrt{\frac{d(d-1)^2}{d+1}}$.
Although this bound is very close to the trivial bound $(d-1)$, we note that better results on the average $l_1$ norm of coherence of random pure states and their typical nature can be obtained \footnote{Private communication with Kaifeng Bu.}.

\section{Random pure quantum states are not typically maximally coherent}\label{sec:max-coh}
It is well known that random bipartite pure states in higher dimension sampled from the uniform Haar measure are maximally entangled with an overwhelmingly large probability \cite{Hayden2006}. Our explorations in previous parts suggest that the randomly chosen pure states are not typically maximally coherent (to within an arbitrarily small error) as they have their relative entropy of coherence concentrated around $H_d-1\neq \ln d$ (see also Fig. \ref{fig}). Here we make this observation precise by proving that indeed the trace distance between the diagonal part of a random pure state and the maximally mixed state does not typically go to zero in the higher dimension case, instead it is almost always concentrated around a fixed nonzero value. To establish this, we use the following lemma.
\begin{lem}
 \label{avg-dis}
Let $\ket{\psi}$ be a random pure state in a $d$ dimensional Hilbert space. The average trace distance between the diagonal part of a random pure state and the maximally mixed state is given by $2(1-1/d)^d$, i.e.,
\begin{align}
\mathbb{E}_\psi \left|\left| \rho_D(\psi) - \frac{\mathbb{I}}{d} \right|\right|_1= 2\left(1-\frac1d\right)^d.\nonumber
\end{align}
\end{lem}
\begin{mproof}
Consider a pure state $\ket{\psi}=\sum^d_{j=1}\psi_j\ket{j}$ with $\psi_j=\bra{j}\psi\rangle=x_j+i y_j$, $i=\sqrt{-1}$ and $x_j, y_j \in\real(j=1,\ldots,d)$. The unique, normalized, unitary invariant measure $\dif (\psi)$ upon the pure state  manifold of normalized state vectors $\ket{\psi}$ is realized by the following delta function prescription 
\begin{eqnarray*}
\frac{\Gamma(d)}{\pi^d}\delta\Pa{1-\sum^d_{j=1}(x^2_j+y^2_j)}\prod^d_{j=1}\dif x_j\dif y_j,
\end{eqnarray*}
if one is interested in calculating the averages of the functions of the form $f(\bra{\psi}\hat P\ket{\psi})$, where $\hat P$ is a projector \cite{Jones1991}. This is the case for us. Here $\Gamma(d)$, which is equal to $(d-1)!$, is the Gamma function.
By performing change of variables, namely, $x_j=\sqrt{r_j}\cos\theta_j$ and $y_j=\sqrt{r_j}\sin\theta_j$ in above for each $j$ with $r_j\geqslant 0$ and $\theta_j\in[0,2\pi]$, $\dif (\psi)$ can also be realized as
\begin{eqnarray*}
\frac{\Gamma(d)}{(2\pi)^d}\delta\Pa{1-\sum^d_{j=1}r_j}\prod^d_{j=1}\dif r_j \dif \theta_j.
\end{eqnarray*}
For a fixed reference basis $\{\ket{j}\} (j=1,\ldots,d)$, we have $\rho_D(\psi)=\sum^d_{j=1}\abs{\psi_j}^2\proj{j}$ with $r_j:=x_j^2+y_j^2=\abs{\psi_j}^2$. Now
\begin{align}
&\mathbb{E}_\psi\norm{\rho_D(\psi)-\frac{\I_d}d}_1\nonumber\\
&= \int \dif (\psi)\Pa{\sum^d_{j=1}\abs{\abs{\psi_j}^2-\frac1d}}\nonumber\\
&= \Gamma(d)\int\Pa{\sum^d_{j=1}\abs{r_j-\frac1d}}\delta\Pa{1-\sum^d_{j=1}r_j}\prod^d_{j=1}\dif r_j \nonumber\\
&= \Gamma(d+1)\int^1_0\dif r_1\abs{r_1-\frac1d}\int^\infty_0\delta\Pa{(1-r_1)-\sum^d_{j=2}r_j}\prod^d_{j=2}\dif r_j\nonumber\\
&=\frac{\Gamma(d+1)}{\Gamma(d-1)}\mathcal{K}.
\end{align}
where $ \mathcal{K}=\int^1_0\dif r_1\abs{r_1-\frac1d}(1-r_1)^{d-2}$. In what follows, we calculate the integral $\mathcal{K}$.
\begin{eqnarray}
\label{int-int}
\mathcal{K}&=&\int^{\frac1d}_0\dif r_1\Pa{\frac1d-r_1}(1-r_1)^{d-2}+\int^1_{\frac1d}\dif r_1\Pa{r_1-\frac1d}(1-r_1)^{d-2}\nonumber\\
&=&\frac{-2}{d(d-1)}\left\{ \left( \frac{d-1}{d} \right)^{d-1} -1 \right\} -2 \int^{\frac1d}_0\dif r_1 r_1(1-r_1)^{d-2}.
\end{eqnarray}
Now
\begin{eqnarray*}
\int^{\frac1d}_0 r_1(1-r_1)^{d-2}\dif r_1 &=& \frac{-1}{d(d-1)}\left[\left( \frac{d-1}{d} \right)^{d-1} + \left( \frac{d-1}{d} \right)^{d}-1\right].
\end{eqnarray*}
Putting above in Eq. (\ref{int-int}), we get $\mathcal{K}= \frac2{d(d-1)}\Pa{1-\frac1d}^d$. Therefore,
\begin{eqnarray}
\mathbb{E}_\psi\norm{\rho_D(\psi)-\frac{\I_d}d}_1
&=&2\Pa{1-\frac1d}^d.
\end{eqnarray}
This completes the proof of the lemma.
\end{mproof}
In the following theorem, we establish that most of the Haar distributed pure quantum states are not typically maximally coherent (within an arbitrarily small error). The main idea is to show that the trace distance of the diagonal part of any random pure quantum state from the maximally mixed state is almost always concentrated around a nonzero number, even in $d\rightarrow\infty$ limit.
\begin{thm}
\label{mx-dis}
Let $\ket{\psi}$ be a random pure state in a $d$ dimensional Hilbert space. The probability that the trace distance between the diagonal part of a random pure state and the maximally mixed state is not close to $ 2\Pa{1-\frac1d}^d$ is bounded from above by an exponentially small number in the large $d$ limit, i.e., for all $\epsilon>0$
\begin{align}
\mathrm{Pr} \left\{ \left|~ \left|\left| \rho_D(\ket{\psi}) - \frac{\mathbb{I}}{d} \right|\right|_1 - 2\Pa{1-\frac1d}^d\right|  > \epsilon \right\}  \leq 2\exp\left( -\frac{d \epsilon^2}{18 \pi^3\ln 2} \right).\nonumber
\end{align}
\end{thm}
\smallskip{}
\begin{mproof}
The Lipschitz constant for the function $F: \ket{\psi} \rightarrow \left|\left| \rho_D (\psi) - \frac{\mathbb{I}}{d} \right|\right|_1$ is $2$ and it can be shown as follows:
\begin{align}
\left|F(\ket{\psi}) - F(\ket{\phi})\right| &= \left|~  \left|\left| \rho_D (\psi) - \frac{\mathbb{I}}{d} \right|\right|_1 -  \left|\left| \rho_D (\phi) - \frac{\mathbb{I}}{d} \right|\right|_1 \right|\nonumber\\
&\leq \left|\left| \rho_D (\psi)  -  \rho_D (\phi) \right|\right|_1 \nonumber\\
&\leq  \left|\left|~ \ket{\psi}\bra{\psi}  - \ket{\phi}\bra{\phi}~ \right|\right|_1 \nonumber\\
&\leq 2 \sqrt{2(1 - \mathrm{Re}(\bra {\psi} \phi \rangle)} = 2  \left|\left| ~\ket{\psi} - \ket{\phi}~\right|\right|,\nonumber
\end{align}
where in the second line we have used the reverse triangle inequality $| || A ||_1 -  || B  ||_1 | \leq || A - B||_1$.  Therefore, $F: \ket{\psi} \rightarrow  \left|\left| \rho_D (\psi) - \frac{\mathbb{I}}{d}  \right|\right|_1$ is a Lipschitz continuous function with the Lipschitz constant $\eta = 2$. Now applying L\'evy's lemma for averages to the function $\left|\left| \rho_D(\ket{\psi}) - \frac{\mathbb{I}}{d} \right|\right|_1$, we obtain
\begin{align}
\mathrm{Pr} \left\{ \left|~ \left|\left| \rho_D(\ket{\psi}) - \frac{\mathbb{I}}{d} \right|\right|_1 - \mathbb{E}_\psi\left|\left| \rho_D(\ket{\psi}) - \frac{\mathbb{I}}{d} \right|\right|_1\right|  > \epsilon \right\}  \leq \eta,\nonumber
\end{align}
where $\eta=2\exp\left( -d \epsilon^2/18 \pi^3\ln 2 \right)$. We complete the proof of the theorem by using Lemma \ref{avg-dis} in the above expression.
\end{mproof}
Theorem \ref{mx-dis} tells us that for majority of pure quantum states the trace distance of the diagonal part from the maximally mixed state is concentrated around $2\Pa{1-\frac1d}^d$, which for $d\rightarrow \infty$ converges to $2/e=0.7357$. Therefore, the diagonal part of most of random pure quantum states maintains a fixed finite distance from the maximally mixed state. Thus, Theorem \ref{mx-dis} implies that the overwhelming majority of random pure quantum states are not typically maximally coherent (within an arbitrarily small error). Next, we find a lower bound on the relative entropy of coherence of the majority of random pure quantum states, for which $\left|\left| \rho_D(\ket{\psi}) - \frac{\mathbb{I}}{d} \right|\right|_1 = 2\Pa{1-\frac1d}^d$. Utilizing the Fannes-Audenaert inequality \cite{Fannes1973, Audenaert2007}, we have
\begin{align}
 \abs{S\Pa{\frac{\mathbb{I}}{d}}- S(\rho_D(\psi))}=\ln d - S(\rho_D(\psi)) &\leq T \ln (d-1) + H_2(T)\nonumber\\
 &\leq T \ln d + H_2(T),\nonumber
\end{align}
where $T =\left|\left| \rho_D(\ket{\psi}) - \frac{\mathbb{I}}{d} \right|\right|_1/2 = \Pa{1-\frac1d}^d$ and $H_{2}(T) = -T \ln T - (1-T)\ln (1-T) $ is the binary Shanon entropy. Therefore,
\begin{align}
\label{fannes-bound2}
 C_{r}(\psi) = S(\rho_D(\psi)) \geq (1-T)\ln d - H_2(T).
\end{align}
Combining Eq. (\ref{fannes-bound2}) with Theorem \ref{mx-dis}, we conclude that the relative entropy of coherence of a randomly picked pure state is, with high probability, always greater than $(1-T)\ln d - H_2(T)$. For $d\rightarrow\infty$, we have
\begin{align}
 \lim_{d\rightarrow\infty}C_{r}(\psi)/\ln d &\geq 1-\frac1e - \lim_{d\rightarrow\infty}H_2(T)/\ln d\nonumber\\
 &=1-\frac{1}{e}\approx 0.6321.
\end{align}

\section{Conclusion and Outlook}\label{sec:conclusion}
In this work we have established various generic aspects of quantum coherence of random pure states sampled from the uniform Haar measure. We have shown that the amount of relative entropy of coherence of a pure state picked randomly with respect to the Haar measure, with a very high probability, is arbitrarily close to the average relative entropy of coherence, which is given by $\sum^d_{k=2} 1/k$ for a $d$-dimensional system.
%
%
In other words, an overwhelming majority of the pure states have coherence equal to the expected value, within an arbitrarily small error.
This also establishes the typical wave nature of a quantum particle in a random pure state. Further, we find a large subspace (of appropriate dimension) of the total Hilbert space of a quantum system such that for every pure state in this subspace, the relative entropy of coherence (also equal to the distillable coherence \cite{Winter2015}) is almost always greater than a fixed number (depending on the dimension of the Hilbert space) that is arbitrarily close to the typical value of coherence. Also, for every state (pure or mixed) in this subspace, the coherence of formation is almost always bounded from below by the same fixed number.
%
Therefore, quantum states in these subspaces can be useful for many coherence consuming protocols.
Further, we find the expected value of classical purity of randomly chosen pure states, which is then used to find an upper bound on the $l_1$ norm of coherence exploiting known relations between coherence and classical purity.
%
%
Furthermore, we find the average distance of the diagonal part of a
randomly chosen pure
quantum state from the maximally mixed state. We show that diagonal
part of most of random pure states maintains a fixed nonzero distance 
from the maximally mixed state,
thus establishing its typicality. This
amounts to stating that most of the randomly chosen pure states are not typically maximally coherent (within an arbitrarily small error).

The results obtained in our work show the strong typicality of measures of coherence and establish that the description of coherence properties of the Haar distributed pure states, in larger dimensions, only requires a small number of typical parameters such as the Hilbert space dimension. These parameters appear in formulation of the concentration of measure phenomenon. This, in turn, reduces a lot the complexity of coherence theory with respect to the Haar distributed pure states. In the future, it will be very interesting, from practical view point, to estimate the dimension of the largest subspace such that it contains no incoherent state, unlike our result, where we find the dimension of the subspace containing at least some fixed nonzero amount of coherence.

\smallskip
\noindent
{\it Acknowledgments:--}We thank the anonymous referees for helpful comments on our manuscript. L.Z. thanks the National Natural Science Foundation of China (No.11301124) for support, and he is also grateful to Zhen-Peng Xu for providing help in numerical checks, and to Hongchao Kang for comments on the integrals of special functions. U.S. thanks Himadri Shekhar Dhar for various helpful discussions and suggestions. U.S. acknowledges the research fellowship of Department of Atomic Energy, Government of India.

\bibliographystyle{apsrev4-1}
 \bibliography{typ-coh-lit}

\end{document}